\documentstyle[aps,prl,epsf]{revtex}

\newcommand{\tg}{\tilde{g}}

\begin{document}
\draft
\twocolumn[\hsize\textwidth\columnwidth\hsize\csname
@twocolumnfalse\endcsname
\title{Resonant-Cavity-Induced Phase Locking and Voltage Steps in a 
Josephson Array} 
\author{E. Almaas\cite{email1} and D. Stroud\cite{email2}}
\address{Department of Physics, The Ohio State University, Columbus,
Ohio 43210}

\date{\today}
\maketitle
\begin{abstract}
We describe a simple dynamical model for an underdamped Josephson
junction array coupled to a resonant cavity.  From numerical solutions
of the model in one dimension, we find that (i) current-voltage
characteristics of the array have self-induced resonant steps (SIRS),
(ii) at fixed disorder and coupling strength, the array locks into a
coherent, periodic state above a critical number of active Josephson
junctions, and (iii) when $N_a$ active junctions are synchronized on
an SIRS, the energy emitted into the resonant cavity is quadratic with
$N_a$.  All three features are in agreement with a recent experiment
\protect{[Barbara {\it et al}, Phys. Rev. Lett. {\bf 82}, 1963
(1999)]}.
\end{abstract}

\pacs{PACS numbers: 05.45.Xt, 79.50.+r, 05.45.-a, 74.40.+k}
\vskip1.5pc] 

Arrays of Josephson junctions have long been studied both
experimentally\cite{micro} and theoretically\cite{theory} as a
potentially controllable source of microwave radiation.  Most studies
have been carried out on overdamped junction arrays with external
loads.  Typically, a dc current is injected into the array, producing
ac voltage oscillations in each of the junctions.  If all the
junctions are locked to the same frequency, then the radiated power
should vary as the square of the number of junctions.  Overdamped
junctions are usually studied, because underdamped junctions can
exhibit hysteresis and chaotic behavior.  However, even overdamped
arrays have proven difficult to synchronize: their largest
experimentally achieved dc to ac conversion efficiency is only about
1\% \cite{yield}.

Recently, Barbara {\it et al} \cite{barbara} achieved a 17\% degree of
power conversion in an {\em underdamped} two-dimensional array placed
within a resonant electromagnetic cavity.  In this case, the
synchronization was achieved by an indirect coupling between each
junction and the electromagnetic field of the cavity mode.  The
results were characterized by striking threshold behavior: typically
no synchronization was achieved for arrays shorter than a certain
threshold number of junctions.

In this Rapid Communication, we present and numerically study a simple
model for the dynamics of an underdamped Josephson junction array
coupled to a resonant cavity.  This model generalizes one used
recently to describe the energetics of such a system \cite{harb}.  It
bears many resemblances to previous dynamical models, which either
connect this array to laser action in excitable two-level
atoms\cite{xx} or introduce various types of impedance loads to
provide global coupling between
junctions\cite{wiesenfeld,cawthorne,filatrella,abdullaev}.  In our
model, we infer the equations of motion starting from a more
conventional Hamiltonian which describes Josephson junctions coupled
to a vector potential\cite{teitel}.  Even though our model is only
one-dimensional, our results show many of the features seen
experimentally\cite{barbara}, including (i) mode locking into a
coherent state above a critical number $N_c$ of active junctions, (ii)
a quadratic dependence of the energy on the number of active junctions
above $N_c$, and (iii) most strikingly, self-induced steps at voltages
corresponding to multiples of the cavity frequency.

We begin with the following Hamiltonian model for a one-dimensional
array of $N$ Josephson junctions placed in a resonant cavity, which we
assume supports only a single photon mode of frequency $\Omega$:
\begin{eqnarray}
H & ~=~& H_{photon} + H_C + H_J \nonumber\\
  & ~=~& \hbar \Omega (a^\dag a+\frac{1}{2}) + \sum_{j=1}^NE_{Cj} n_j^2 - 
	\sum_{j=1}^NE_{Jj} \cos (\gamma_j) \label{eq:h}.
\end{eqnarray}
Here, $H_{photon}$ is the energy in the cavity, $H_C$ is the
capacitive energy, and $H_J$ is the Josephson energy of the
array. $a^\dag$ and $a$ are photon creation and annihilation
operators, $E_{Cj} = q^2/(2C_j)$ is the approximate capacitive energy,
and $E_{Jj} = \hbar I_{cj}/q$ is the Josephson coupling energy of a
junction (where $C_j$ is a capacitance, $I_{cj}$ a critical current,
and $q = 2|e|$ is the Cooper pair charge).  Finally, $\gamma_j =
\phi_j - [(2\pi)/\Phi_0] \int_j {\bf A} \cdot {\bf ds} \equiv~ \phi_j -
A_j$ is the gauge-invariant phase difference across a junction, where
$\phi_j$ is the phase difference across a junction in the absence of
the vector potential ${\bf A}$, $\Phi_0=hc/q$ is the flux quantum, and
the line integral is taken across the junction.  We assume that ${\bf
A}$ arises from the electromagnetic field of the normal mode of the
cavity.  In Gaussian units, it is given by\cite{slater,yariv} ${\bf
A}({\bf x},t) = \sqrt{(h c^2)/(2 \Omega)}\left(a(t) +
a^\dag(t)\right){\bf E}({\bf x})$, where ${\bf E}({\bf x})$ is the
electric field of the mode, normalized such that $\int_Vd^3x|{\bf
E}({\bf x})|^2 = 1$, $V$ being the system volume.
Similarly, \cite{yariv} $A_j ~=~ \sqrt{g_j}(a + a^\dag)$, where
\begin{equation}
g_j = \frac{\hbar c^2}{2\Omega} \frac{(2\pi)^{3}}{ 
	\Phi_0^2} \left[\int_j{\bf E}({\bf x})\cdot{\bf ds}\right]^2
\end{equation}
is the effective coupling to the cavity. 

The time-dependence of the operators $a$, $a^\dag$, $n_i$, and
$\phi_i$ follows from Eq.\ (\ref{eq:h}) together with the Heisenberg
equations of motion $i\hbar\dot{{\cal O}} = [{\cal O}, H]$, where
$[..,..]$ is a commutator and ${\cal O}$ an operator.  We use
$[a,a^\dag]=1$, $[a,a]=[a^\dag,a^\dag]=0$, and $[n_j,\pm \phi_k]= \mp
i \delta_{jk}$, with other commutators equal to zero, and the relation
$[A,F(B)]=[A,B]F'(B)$.

We introduce the notation $a = a_R + i a_I$, $\omega_{pj}^2 = 2
\omega_{Cj} \omega_{Jj}$, where $\omega_{Cj} = E_{Cj}/\hbar$ and
$\omega_{Jj} = E_{Jj}/\hbar$, and a dimensionless natural time $\tau =
\bar{\omega}_pt$, with $\bar{\omega}_p$ as a suitable average value of
$\omega_{pj}$.  For numerical convenience, we also assume that $g_j$
has the same value $g$ for each junction.  Then the equations of
motion can be written (in properly scaled units) as $\dot{\phi}_j -
\tilde{n}_j = 0$, $\dot{\tilde{n}}_j + (\omega_{pj}^2 /
\bar{\omega}_p^2) \sin\left(\phi_j-2\tilde{a}_R\right)= 0$,
$\dot{\tilde{a}}_R -\tilde{\Omega}\tilde{a}_I =0$, and
$\dot{\tilde{a}}_I + \tilde{\Omega} \tilde{a}_R - g \sum_j
(\omega_{Jj} / \bar{\omega}_p) \sin \left( \phi_j - 2 \tilde{a}_R
\right) = 0$, where the dot is a derivative with respect to $\tau$.

In order to make these exact relations amenable to numerical
computation, we now replace the operators by $c$-numbers, as should be
reasonable when the eigenvalues of $n_j \gg 1$\cite{xx}.  To introduce
dissipation into the equations of motion, we may add a term to $H$ of
the form $\sum_{j=1}^N\left[\phi_j \sum_\alpha f_\alpha^{(j)}
x_\alpha^{(j)} + \sum_\alpha \left( \frac{1}{2m_\alpha}
(p_\alpha^{(j)})^2 + \frac{m_\alpha}{2} \omega_\alpha^2
x_\alpha^{(j)^2} \right) \right]$, where the $f_\alpha^{(j)}$ are
random variables and $x_{\alpha}^{(j)}$ and $p_\alpha^{(j)}$ are
canonically conjugate\cite{chakra}.  If the spectral density $J_j
\equiv(\frac{\pi}{2})\sum_\alpha\frac{(f_\alpha^{(j)})^2} {m_\alpha
\omega_\alpha} \delta ( \omega - \omega_\alpha ) = \frac{\hbar}{2\pi}
\alpha_j| \omega| \theta ( \omega_c - \omega )$, where $\omega_c$ is a
cutoff frequency comparable to a typical phonon frequency, $\alpha_j =
R_0 / R_j$, and $R_0 = h/(2e)^2$, then the dissipation is
ohmic\cite{leggett} and integrating out the variables $x_\alpha^{(j)}$
and $p_\alpha^{(j)}$ leads to the usual resistively-shunted junction
equation\cite{tinkham} with ohmic damping corresponding to a shunt
resistance $R_j$. A driving current current can be included similarly
by adding to H a ``washboard potential'' of the form $\frac{\hbar
I}{q}\sum_{j=1}^N\phi_j$.  These modifications lead to the following
equations of motion for the $2N+2$ variables:
\begin{equation}\left.
\begin{tabular}{rcl}
$\dot{\phi}_j$      &=& $\tilde{n}_j$, \nonumber\\
$\dot{\tilde{n}}_j$ &=& $\frac{I}{I_c(1+\Delta_j)} -
	\frac{1}{Q_J} \tilde{n}_j - \sin\left(\phi_j - 2 
	\tilde{a}_R\right)$, \nonumber \\
$\dot{\tilde{a}}_R$ &=& $ \tilde{\Omega}\tilde{a}_I$, \nonumber \\
$\dot{\tilde{a}}_I$ &=& $-\tilde{\Omega}  \tilde{a}_R + \tg \sum_j 
	(1+\Delta_j) \sin\left(\phi_j -2 \tilde{a}_R \right)$.
\end{tabular}\right\} \label{eq:1}
\end{equation}
Here, we have redefined the effective coupling as $\tg=g \omega_J /
\bar{\omega}_p$ , and introduced a damping coefficient $Q_J =
\bar{\omega}_p R_jC_j$, where $R_j$ is the shunt resistance. We also
introduce a disorder parameter $\Delta_j = (I_{cj}-I_c)/I_c$, where
$I_c$ is a suitable average critical current. In writing these
equations, we have assumed that both $C_j R_j$ and $I_{cj}/C_j$ are
independent of $j$, so that each junction has the same damping
coefficient $Q_J$.  Dissipation due to the cavity walls could be
included similarly via a cavity Q factor. Note that the first two
equations in (\ref{eq:1}) reduce to the RCSJ model in the limit of no
coupling to the cavity ($\tg = 0$), and the last two equations to
those of a harmonic oscillator with eigenfrequency $\tilde{\Omega}$.

We have solved Eqs.\ (\ref{eq:1}) for the variables $n_i$, $\phi_i$,
$\tilde{a}_R$ and $\tilde{a}_I$ numerically by implementing the
adaptive Bulrisch-Stoer method, which is both fast and accurate
\cite{numrec}.  We choose $I_{cj}$, for each junction, $j$, randomly
and independently from a uniform distribution between $I_c(1-\Delta)$
and 
\begin{figure}[tb]
\epsfysize=7cm
\centerline{\epsffile{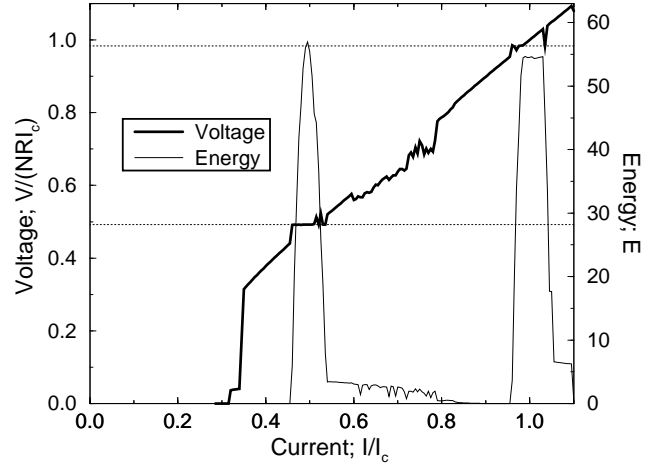}}
\caption{Left scale: Current-voltage (IV) characteristics for an
         underdamped Josephson array of $N=40$ junctions and system
         parameters $\tilde{\Omega} = 2.2$, $Q_J = \sqrt{20}$, $\Delta
         = 0.05$, and $\tg = 0.001$, as defined in the text.
         Right-hand scale: total photon energy in the cavity
         $\tilde{E} = (\tilde{a}_R^2 + \tilde{a}_I^2)$. Predicted
         voltages for the integer self-induced resonant steps (SIRS)
         are shown as dotted lines.}
\label{fig1}
\end{figure}
\noindent
$I_c(1+\Delta)$, but for convenience assume that $\tg$ is
independent of $j$.  We initialize the simulations with all the phases
randomized between $[0,2\pi]$, and $a_R = a_I = n_j = 0$. We then let
the system equilibrate for a time interval $\Delta\tau=10^4$, after
which we evaluate averages over a time interval $\Delta\tau=2\cdot
10^3$, using $2^{16}$ evenly spaced sampling points.

In Fig.\ \ref{fig1}, we show the current-voltage characteristics for
$N = 40$ junctions with $\Delta=0.05$ and $\tg = 0.001$, evaluating
the time-averaged voltage from the Josephson relation, $\langle V
\rangle = [1/Q_J] \langle \sum_{j=1}^N \dot{\gamma}_j\rangle$.  A
striking feature of this plot is the {\em self-induced resonant steps}
(SIRS), at which $\langle V \rangle$ remains approximately constant
over a range of applied current.  The most prominent step occurs at
$\langle V \rangle/(NRI_c) = \tilde{\Omega}/Q_J$, but there is
another, less obvious, step at $2\tilde{\Omega}/Q_J$.  We believe the
steps occur at all $(m/n)\tilde{\Omega}/Q_J$, where $m$ and $n$ are
integers, as further discussed below.  Similar steps were seen
experimentally in a {\em two-dimensional} array of underdamped
Josephson junctions coupled to a resonant cavity\cite{barbara}.  As
noted in Ref.\ \cite{barbara}, these steps are the analog of Shapiro
steps in conventional Josephson junctions.  They occur, we believe,
for a similar reason: qualitatively, the variables $a_R$ oscillate
with frequency $\tilde{\Omega}$ and produce an effective ac drive to
each Josephson junction, in addition to the dc drive generated by the
current $I$.

When we solve the system of equations (\ref{eq:1}) numerically for a
single junction, we find SIRS for fractions $(n/m) = 1, 4/3, 3/2, 5/3,
2, 5/2, 3, 4, ... $.  The step width in current is very sensitive to
$\tilde{g}$, and, indeed, we have thus far found the steps only for a
limited range of $\tilde{g}$.  For the larger arrays, we have not yet
seen the fractional SIRS.

\begin{figure}[tb]
\epsfysize=7cm
\centerline{\epsffile{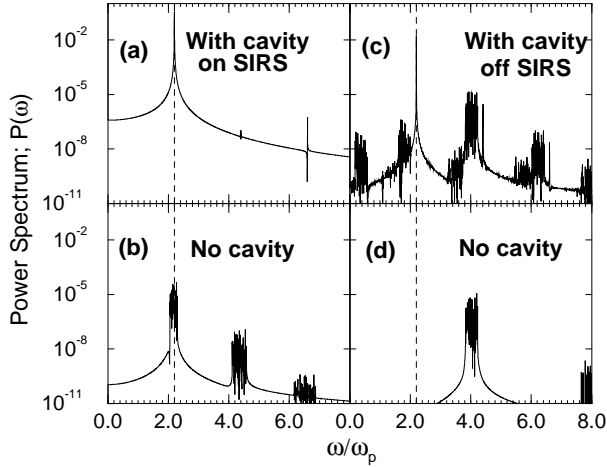}}
\caption{Power spectrum, $P(\omega)$, of the ac voltage across the
         array, plotted versus frequency, $\omega$, at two driving
         currents: (a) and (b) $I/I_c = 0.492$, corresponding to the
         first integer SIRS, and (c) and (d) $I/I_c =0.90$, slightly
         off a SIRS.  Other parameters are the same as in Fig.\
         \ref{fig1}.  (b) and (d) are the same as (a) and (c) except
         that $\tg = 0$.  The vertical, dashed line shows the resonant
         frequency of the cavity.}
\label{fig2}
\end{figure}

Fig.\ \ref{fig1} also shows the time-averaged total energy $\tilde{E}
= (\tilde{a}_R^2 + \tilde{a}_I^2)$ as a function of $I/I_c$ for the
same array.  The total energy in the cavity increases dramatically
when the array is on a SIRS, and is very small otherwise.  This sharp
increase signals the onset of coherence within the array.

Figure\ \ref{fig2} shows the calculated voltage power spectrum
$P(\omega) = |\int_{-\infty}^\infty V_{tot} (\tau) \exp(i\omega\tau)
d\tau |^2$, for two values of the driving current: $I/I_c =
\tilde{\Omega}/Q_J$ [Fig.\ \ref{fig2}(a) and (b)] and $I/I_c = 0.9$
[Fig.\ \ref{fig2}(c) and (d)]; all other parameters are the same as in
Fig. \ref{fig1}.  In (a), all the junctions are on the first SIRS and
the power spectrum has peaks at the cavity frequency $\tilde{\Omega}$
and its harmonics.  In (c), the array is tuned off the step.  The
power spectrum shows that the array is not synchronized in this case;
instead, the individual junctions oscillate approximately at their
individual resonant frequencies and their harmonics and subharmonics.
In Fig.\ \ref{fig2}(b) and (d), we show the same case as in Fig.\
\ref{fig2}(a) and (c) respectively, except that the coupling constant,
$\tg$, is artificially set to zero.  In this case, the junctions are,
of course, independent of one another, and the result is that of a
disordered one-dimensional Josephson array with no coupling between
the junctions.

Next, we turn to the dependence of these properties on the {\em number
of active junctions}, $N_a$, in the array.  The concept of active
junction number, in the terminology of Ref.\ \cite{filatrella}, is
meaningful only for underdamped junctions.  As is well known, an
underdamped junction is bistable and hysteretic in certain ranges of
current, and can have either zero or a finite time-averaged voltage
across it, depending on the initial conditions.  In the present case,
\begin{figure}[tb]
\epsfysize=7cm
\centerline{\epsffile{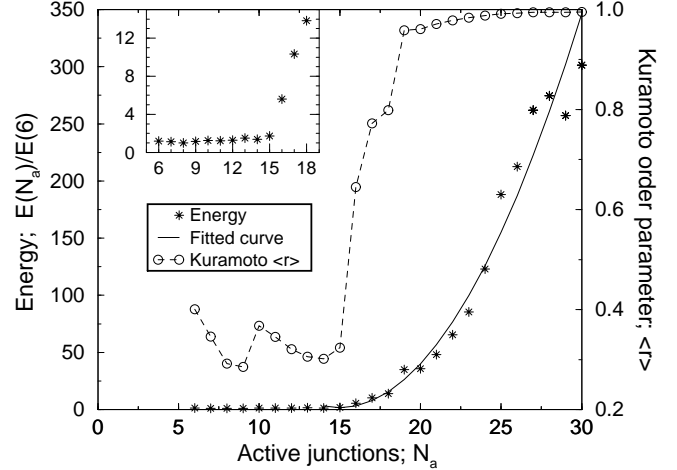}}
\caption{Left-hand scale and asterisks: Photon energy $\tilde{E}$ in
         the resonant cavity when the array is current driven on a
         SIRS, plotted versus number of active junctions, $N_a$.  The
         array parameters are $N=40$, $\tilde{\Omega} = 2.2$, $Q_J =
         \sqrt{20}$, $\Delta = 0.10$, $\tg = 0.001$, and $I/I_c =
         \tilde{\Omega}/Q_J$ (see text).  Full curve shows the best
         fit of $\tilde{E}$ to the function $c_2 N_a^2 + c_1 N_a +
         c_0$ for $N_a > 15$, the threshold for synchronization.
         Inset: $\tilde{E}(N_a)$ near $N_a = 15$, showing jump near
         synchronization threshold.  Right-hand scale and open
         circles: Kuramoto order parameter, $\langle r \rangle$ (see
         text), for the same array.  Dots connecting circles are
         guides to the eye.  The sharp increase in $\langle r \rangle$
         and the quadratic increase in $\tilde{E}$ both start near
         $N_a = 15$.}
\label{fig3}
\end{figure}
\noindent
$N_a$ denotes the number of junctions (out of $N$ total) which have a
finite time-averaged voltage drop.  We can tune $N_a$ by suitable
choosing the initial conditions, $\phi_i$ and $\dot{\phi}_i$, for each
junction\cite{filatrella}.

We have studied the properties of the disordered array ($\Delta =
0.10$) for $N = 40$ junctions, and a driving current $I/I_c =
\tilde{\Omega}/Q_J$.  This current not only lies well within the
bistable region, but also leads to a voltage on the first integer
SIRS.  The total energy $\tilde{E}(N_a)$ (normalized to
$\tilde{E}(6)$) for this case is plotted as a function of $N_a$ in
Fig.\ \ref{fig3}.  The active junctions are unsynchronized up to a
threshold value $N_c = 15$.  Above this value, $\tilde{E}$ increases
as a quadratic function of $N_a$, i.\ e., $\tilde{E} = c_0 + c_1 N_a +
c_2 N_a^2$, where $c_0$, $c_1$, and $c_2$ are constants (full line in
Fig.\ \ref{fig3}).  By contrast, $\tilde{E}$ is approximately
independent of $N_a$ for $N_a < N_c$.  At $N_a = N_c$, there is a
discontinuous jump in $\tilde{E}$ by approximately a factor of $3$
(see inset to figure).  A similar quadratic dependence above a
synchronization threshold was also seen in Ref.\ \cite{barbara},
though for a two-dimensional array in an applied weak magnetic field.
By contrast, if the system is in the bistable region, but {\em not}
tuned to a self-induced resonant step, $\tilde{E}$ does {\em not}
increase quadratically with $N_a$. Instead, we find $\tilde{E}(N_a)$
exhibits a series of plateaus separated by discontinuous jumps (not
shown in the Figure).

To measure the degree of synchronization among the Josephson
junctions, we plot the {\em Kuramoto order parameter} \cite{kuramoto},
$\langle r \rangle$ for the same parameters, as a function of number
of active junctions, $N_a$, (right-hand scale in Fig.\ \ref{fig3}).
$\langle r \rangle$ is defined by $\langle r \rangle = \langle |
\frac{1}{N_a} \sum_{j=1}^{N_a} \exp(i\phi_j)| \rangle_\tau$, where
$\langle...\rangle_\tau$ denotes a time average.  Note that $\langle r
\rangle = 1$ represents perfect synchronization, while $\langle r
\rangle = 0$ would correspond to no correlations between the different
phase differences $\phi_i$.  As is clear from Fig.\ \ref{fig3}, there
is an abrupt increase in $\langle r \rangle$ at $N_a = N_c$,
indicative of a {\em dynamical transition} from an unsynchronized to a
synchronized state, as $N_a$ is increased past a critical value, while
other parameters are kept fixed.  As expected from similar transitions
in other models\cite{strogatz}, the finite this transition is not
inhibited by the finite disorder in the $I_c$'s.  Note that $\langle r
\rangle$ approaches unity for large $N_a$, representing perfect
synchronization.  This transition is the dynamic analog of that
analyzed by an equilibrium mean-field theory in Ref.\ \cite{harb}.
The existence of this transition is intuitively reasonable, as
discussed there: since each junction is effectively coupled to every
other junction via the cavity, the strength of the coupling increases
with $N_a$, and a transition to coherence should occur for
sufficiently large $N_a$.

In summary, we have presented a model for a one-dimensional array of
underdamped Josephson junctions coupled to a resonant cavity.  We have
studied the classical limit of the Heisenberg equations of motion for
this model, valid in the limit of large numbers of photons, and
included damping by coupling each phase difference to an ohmic heat
bath.  In the presence of a dc current drive, we find numerically that
(i) the array exhibits self-induced resonant steps (SIRS), similar to
Shapiro steps in conventional arrays; (ii) there is a transition
between an unsynchronized and a synchronized state as the number of
active junctions is increased while other parameters are held fixed;
and (iii) when the array is biased on the first integer SIRS, the
total energy increases quadratically with number of active junctions.
All these features appear consistent with experiment\cite{barbara}.
Further study is underway in order to ascertain whether or not these
features remain true of two-dimensional arrays and with
gauge-invariant damping.

We are grateful for support from NSF grant DMR97-31511. Computational
support was provided by the Ohio Supercomputer Center, and the
Norwegian University of Science and Technology (NTNU).  We thank C.\
J.\ Lobb and R.\ V.\ Kulkarni for useful conversations.

\end{document}